\documentclass[12pt]{iopart}
\usepackage{graphicx}% Include figure files
\usepackage{dcolumn}% Align table columns on decimal point
\usepackage{bm}% bold math
\usepackage{epsfig}

\def\beq{\begin{equation}}
\def\eeq{\end{equation}}
\def\bea{\begin{eqnarray}}
\def\eea{\end{eqnarray}}
\def\nn{\nonumber}

\newfont{\cursive}{pzcmi at 9pt}

\begin{document}
\parskip 3pt

\title{Stable gravastars with generalised exteriors}

\author{Benedict M.N. Carter}
\eads{\mailto{bmc55@student.canterbury.ac.nz}}
\address{Department of Physics \& Astronomy, University of Canterbury,
Private Bag 4800, Christchurch, New Zealand}
\date{\today}

\\
\begin{abstract}New spherically symmetric gravastar solutions, stable to radial perturbations, are found by utilising the construction of Visser and Wiltshire. The solutions possess an anti--de Sitter or de Sitter interior and a Schwarzschild--(anti)--de Sitter or Reissner--Nordstr\"{o}m exterior. We find a wide range of parameters which allow stable gravastar solutions, and present the different qualitative behaviours of the equation of state for these parameters.
\end{abstract}

\maketitle

\section{Introduction}
Black holes are widely accepted as physical objects due to their mathematical elegance and the strong astronomical evidence for their existence. The study of the properties of black hole horizons is of fundamental importance as horizons appear to provide a strong link between gravitation, thermodynamics and quantum theory. However, these horizons when described semi-classically give rise to a number of seemingly paradoxical theoretical problems which have yet to be satisfactorily resolved (see \cite{waldreview} for a review). For example, one such topic of current interest is whether a pure quantum state which passes over the event horizon of a black hole can evolve into a mixed state during black hole evaporation. This problem is known as the `black hole information paradox'. It is generally accepted that the final resolution of this issue, and other difficult theoretical problems (such as the `blue shift catastrophe' \cite{BirDav}) caused by the current description of black hole horizons, will be achieved using quantum gravity. However, we do not yet have a full theory of quantum gravity, and therefore, other possible solutions to remove the paradoxes generated by black hole horizons should be investigated.

Given the above, it has been suggested that alternative endpoints of gravitational collapse of a massive star, which do not involve horizons, should be studied. The general idea is to prevent the possibility of a horizon forming, by stopping the collapse of matter at some radius greater than that of the horizon. This prevention of the formation of a horizon thereby precludes problems like the black hole information paradox. There are a variety of proposals including the Mazur--Mottola ``gravastar'' (\textbf{gra}vitational \textbf{va}cuum \textbf{star}) \cite{gravastar}. Another scenario envisages the horizon as an emergent property of a Bose superfluid \cite{Chapline:2000en}. Mazur and Mottola's gravastar scenario is a solution of Einstein's equations which has a Schwarzschild exterior, a de Sitter interior corresponding to some sort of Bose condensate, and a rigid spherical shell of matter whose thickness is of the order of the Planck length, suspended approximately a couple of Planck lengths outside of the Schwarzschild radius. Due to their extreme compactness, it seems difficult to observationally distinguish gravastars from black holes. It has been argued that any star, with a surface, will emit much more radiation during accretion than black holes (which have no surface) (see \cite{Done:2002yv} for a review), but it has also been shown that gravastars may be just as `black' as black holes \cite{marek}.

Although the gravastar model forwarded by Mazur and Mottola has not gained much attention by the majority of general relativists, it has led others to construct similar models. Bili\'{c} \textit{et al.} consider a gravastar with a Born-Infeld-phantom interior geometry \cite{Bilic:2005sn}, while Cattoen \textit{et al.} generate a method for creating generalised gravastars with anisotropic continuous pressures \cite{Cattoen:2005he}. 

Visser and Wiltshire \cite{Visser:2003ge} sought to determine whether the Mazur-Mottola gravastar is dynamically stable against radial perturbations. To do so, rather than considering the original Mazur--Mottola gravastar, they considered a simplified model with three layers:
\begin{itemize}
\item{An external Schwarzschild vacuum, $\rho=0=P$}
\item{A single thin shell \cite{shell}, with surface density $\sigma$ and surface tension $\theta$; with radius $a\geq2M$.}
\item{A de Sitter (dS) interior, $P=-\rho$.}
\end{itemize}
Using this model they found a condition for stability for the thin shell against radial perturbations in terms of an `effective energy equation' for a non-relativistic particle with 'energy' $E$ in a 'potential' $V(a)$,
\beq
\frac{1}{2}\dot{a}^2+V(a)=E\,.
\label{energyeq}
\eeq 
See section 4 from \cite{Visser:2003ge} for a precise definition of $V(a)$ and the quantities therein. The thin shell will be stable against radial perturbations if and only if there exists some $a_0$ such that $V(a)$ satisfies
\beq V(a_0)=0; \qquad V'(a_0)=0; \qquad V''(a_0) \geq 0
\eeq
Using Visser and Wiltshire's method one is able to prescribe the interior matter, $m_-(a)$, exterior matter, $m_+(a)$, and the potential $V(a)$ of the gravastar to parametrically find the equation of state for the thin shell. Indeed, as we note below, the mass of the thin shell, $m_s(a) \equiv 4\pi \sigma(a)\;a^2$, can be
calculated as an explicit function of $m_+(a)$, $m_-(a)$ and $V(a)$. 

Visser and Wiltshire demonstrated that there exist large classes of potentials, and consequently, equations of state, for which gravastars are stable against spherically symmetric gravitational perturbations as well as large classes of potentials which are unstable. Consequently, particular choices of potentials and mass functions need to be studied on a case by case basis, since more general criteria for stability are not presently known.

%Visser and Wiltshire analysed, in detail, the stability of a gravastar which has de Sitter interior and Schwarzschild exterior with ``$V(a)\equiv0$''. They found that, given some fine tuning of the interior de Sitter curvature to the exterior mass, stable gravastars which satisfy the dominant energy condition can exist.

In this paper we take the method as outlined by Visser and Wiltshire and analyse the stability of gravastars that have a Schwarzschild-(anti)-de Sitter or Reissner--Norstr\"{o}m exterior. Given our current understanding, we pick these two types of exterior metrics as they are physically reasonable, spherically symmetric and static. 

\section{Schwarzschild--(A)dS Gravastar}
\subsection{Definitions}
Following directly on from the work of Visser and Wiltshire  \cite{Visser:2003ge}, we start with the general equations, (47) and (49), from their paper. These relate the surface density, $\sigma(a)$, and surface tension, $\theta(a)$, of the thin shell which makes up the surface of the gravastar, to the potential and mass functions via
\beq
\sigma(a)=-\frac{1}{4\pi a} \left[\left[ \sqrt{1-2V(a)-\frac{2m(a)}{a}}\right]\right],
\label{sigma}
\eeq
\beq
\theta(a)=-\frac{1}{8\pi a}\left[\left[\frac{1-2V(a)-m(a)/a-aV'(a)- m'(a)}{\sqrt{1-2V(a)-2m(a)/a}}\right]\right].
\label{theta}
\eeq
where $m_-(a)$ is the interior mass profile, $m_+(a)$ is the exterior mass profile, $V(a)$ is the effective energy potential, and
\beq
[[X]]\equiv X_+-X_-\,.
\eeq
See \cite{visser} for more details on this notation. See \cite{Visser:2003ge} for more discussion on the above expressions and definitions. We now specialise the exterior geometry to Schwarzschild-(A)dS, the interior to (A)dS, and the potential to zero. We therefore write
\bea
m_+(a)&=&M-\Lambda a^3/6\,,
\\
m_-(a)&=&ka^3\,,
\\
V(a)&=&0\,,
\eea
where $\Lambda$ is the asymptotic constant spatial curvature of the exterior geometry and $k$ is the curvature of the interior geometry due to a vacuum energy. 
We note that in this analysis, for $E = 0$ in Eq.(\ref{energyeq}), the choice of $V(a) = 0$ provides a stable equilibrium for the gravastar thin shell, as $\dot{a} = 0$, and hence the radius, $a$, cannot change to some other value. Simultaneously, $V(a)=V'(a)=V''(a)=0$ for all values of $a$. This specialization to $V(a)=0$ closely follows the calculation of \cite{Visser:2003ge}. We now convert all variables to dimensionless parameters, parameterised by $M$;
\beq
A=\frac{a}{M}
\,,\
L=\Lambda M^2
\,,\
K=k M^2
\,,\
\rho=\sigma M
\,,\
P=\theta M\,.
\label{dimensionless}
\eeq
This leads to
\beq
\rho(A)=\frac{1}{4\pi A} \left[\sqrt{1-2K A^2}-\sqrt{1-\frac{2}{A}+\frac{L A^2}{3}}\right]\,,
\eeq
\beq
P(A)=\frac{1}{8\pi A}\left[\frac{1-4K A^2}{\sqrt{1-2K A^2}}- \frac{1-\frac{1}{A}+\frac{2L A^2}{3}}{\sqrt{1-\frac{2}{A}+\frac{L A^2}{3}}}\right]\,.
\eeq
Reality of the reparameterised surface density, $\rho$, and surface tension, $P$, requires
\bea
K &<& \frac{1}{2A^2}\,,
\nn
\\
L &>& \frac{6}{A^3}-\frac{3}{A^2}\,.
\label{Lr}
\eea

\subsection{Solutions}
%------------------------------------------------------------
\begin{figure}[hb]
\vbox{\centerline{\scalebox{0.33}{\includegraphics[angle=0]{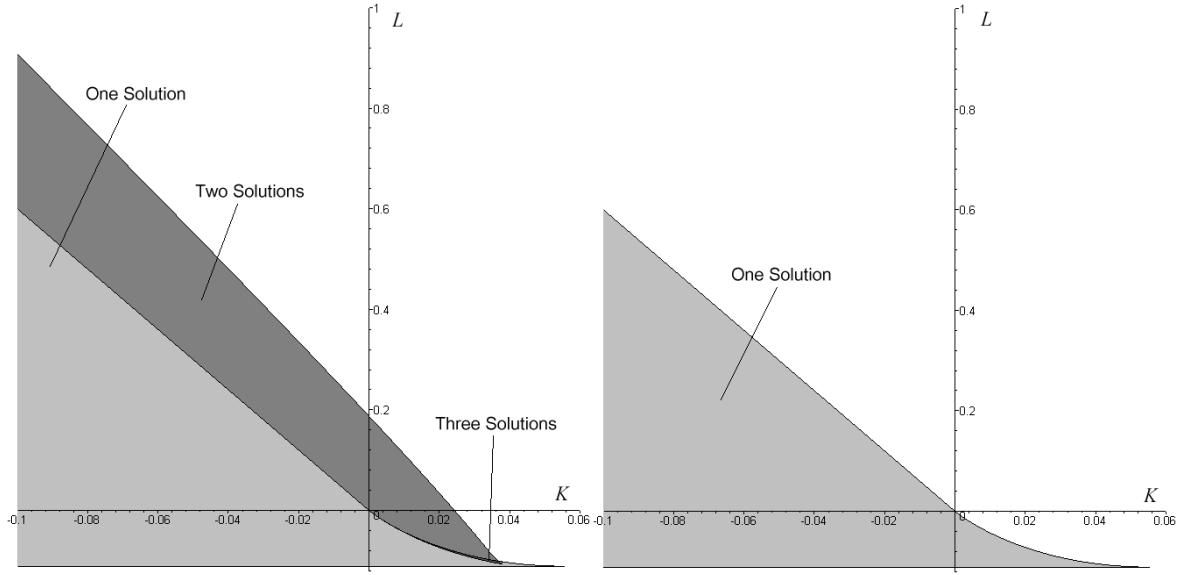}}}
\caption{\label{fig:gravastarLnp}
{\sl The solution space for stable gravastars with (A)dS interior and Schwarzschild--(A)dS exterior satisfying $\rho+P=0$ in the left panel, or $\rho-P=0$ in the right panel. The number of solutions in each graph indicates how many times the equation of state intersects the part of $\mathcal{\partial M}$ that the equation defining the graph represents. The unshaded parts have no solutions, and correspondingly have no interval of $\rho(P)$ which lies in $\mathcal{M}$.}}}
\end{figure}
\begin{figure}[hb]
\vbox{\centerline{\scalebox{0.33}{\includegraphics[angle=0]{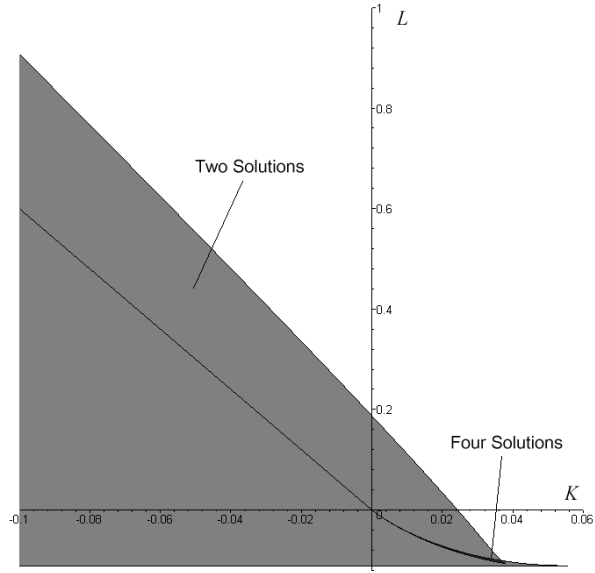}}}
\caption{\label{fig:gravastarL}
{\sl The solution space for stable Gravastars with (A)dS interior and (A)dS Schwarzschild exterior satisfying $\rho+P=0$ or $\rho-P=0$. The unshaded region has no solutions where $\rho(P)$ intersects $\mathcal{\partial M}$, and due to continuity of the the equation of state, no interval of $\rho(P)$ lies in $\mathcal{M}$. The shaded regions with two solutions have one interval of $\rho(P)$ which lies in $\mathcal{M}$, while the region with four solutions has two intervals of $\rho(P)$ satisfying the dominant energy condition.}}}
\end{figure}

%------------------------------------------------------------
We first present a detailed analysis of the dominant energy condition, as it provides the most restrictive conditions on the matter of the thin shell of the gravastar. %Given that the matter is presumed to form through a quantum phase transition, we cannot specifically state the type of matter that forms the thin shell, but as we expect it to be conventional matter we will require it to satisfy the dominant energy condition. 
Given that the matter is presumed to form through a quantum phase transition, we cannot specifically state the type of matter that forms the thin shell, but we shall assume the dominant energy condition continues to hold.
For brevity we define the region satisfying the dominant energy condition by $\mathcal{M}$, and the boundary of $\mathcal{M}$ by $\mathcal{\partial M}$. The left and right sides of $\mathcal{\partial M}$ are defined by the equations
\beq
\rho+P=0\,,\  \rho-P=0\,,
\label{DECs}
\eeq
respectively, satisfying the condition that $\rho \geq 0$. Solutions to the above equations give the points where the equation of state, $\rho(P)$, enters or exits $\mathcal{M}$. After some manipulation Eqs.(\ref{DECs}) become,
\beq
\left(3-\frac{5}{A}+\frac{4L A^2}{3}\right)\sqrt{1-2KA^2} = (3-8KA^2)\sqrt{1-\frac{2}{A}+\frac{LA^2}{3}}\,,
\label{DECLn}
\eeq
and
\beq
(A-3)\sqrt{1-2KA^2} = A\sqrt{1-\frac{2}{A}+\frac{LA^2}{3}}\,,
\label{DECLp}
\eeq
respectively. Squaring these two expressions can lead to an inclusion of a relative minus sign between the different sides of the equations, which we account for by placing extra restrictions on $K$ and $L$. For $K$ and $L$ satisfying (\ref{Lr}) we can see that the \textit{l.h.s.\ }of (\ref{DECLn}) is positive. Therefore the \textit{r.h.s.\ }is also positive, thereby providing the bound $K<\frac{3}{8A^2}$ which applies to Eq.(\ref{DECLn2}) below. Similarly the \textit{r.h.s.\ }of (\ref{DECLp}) is positive, which implies the \textit{l.h.s.\ }is positive and hence $A>3$ for Eq.(\ref{DECLp2}) below. After some further manipulation we find the multivariate polynomials,
\bea
0 &=& \left( -32K{L}^{2}-192L{K}^{2} \right) {A}^{8}+ \left( 16{L}^{2}-576{K}^{2} \right) {A}^{6}
\nn
\\
&\ &+\left( 240KL+1152{K}^{2} \right) {A}^{5}+ \left( 45L+270K \right) {A}^{4}
\nn
\\
&\ &
+ \left( -324K-120L\right) {A}^{3}-450K{A}^{2}-108A+225\,,
\label{DECLn2}
\eea
and,
\beq
0= -12A+27-6KA^4+36KA^3-54KA^2-LA^4\,.
\label{DECLp2}
\eeq
The octic equation in $A$, (\ref{DECLn2}), is not analytically solvable in general. We therefore solve this, and the quartic in $A$, (\ref{DECLp2}), numerically and create a contour plot representing the number of solutions for $A$ in the phase space parameterised by $K$ and $L$. Such solutions will have a number of restrictions placed on them. Specifically, they must satisfy the inequality (\ref{Lr}), $\rho>0$, and any other conditions required for the consistency of (\ref{DECLn}) and (\ref{DECLp}). Figure~\ref{fig:gravastarLnp} displays the number of times the equation of state, $\rho(P)$, crosses the boundaries of $\mathcal{M}$, $\rho+P=0$ and $\rho-P=0$, separately. Figure~\ref{fig:gravastarL} shows the combined results of the two panels in Fig.~\ref{fig:gravastarLnp}. Figure~\ref{fig:gravastarLnp} is useful for determining the qualitative behavior of the equation of state, while Fig.~\ref{fig:gravastarL} is useful for determining the existence of gravastar solutions which have a thin-shell satisfying the dominant energy condition.

%------------------------------------------------------------
%------------------------------------------------------------

One can see from Fig.~\ref{fig:gravastarL} that there are many different regions of $K$ and $L$ that have real and finite solutions to the equations that define $\mathcal{\partial M}$. This means that for those values of $K$ and $L$ there is always an interval of the equation of state, $\rho(P)$, which satisfies the dominant energy condition. This means there can always be gravastars with thin shells of matter which satisfy the dominant energy condition and are stable to radial perturbations. Visser and Wiltshire's gravastar with an exterior Schwarzschild solution is reproduced as a special case in this diagram when $L=0$. The unshaded region has no solutions, implying that the equation of state does not enter or exit $\mathcal{M}$. One can check that for those parameter ranges $\rho(P)$ does not lie in $\mathcal{M}$. This means that for those values of the parameters $K$ and $L$, it is not possible to form a gravastar which has a static thin-shell satisfying the dominant energy condition which is stable under radial perturbations.  

%------------------------------------------------------------
\begin{figure}[h]
\vbox{\centerline{\scalebox{0.50}{\includegraphics[angle=0]{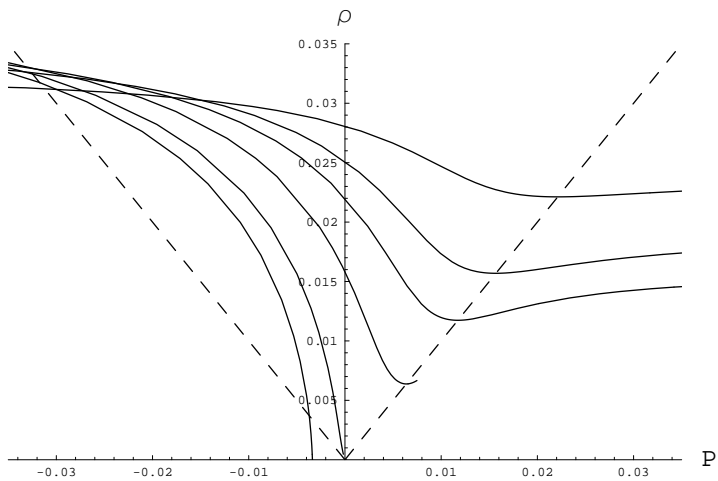}}}
\caption{\label{fig:Ln001}
{\sl The behaviour of $\rho(P)$ for the thin shell of a gravastar with $K=-0.01$ and a range of values of $L$. From left to right, $L=\{0.1,0.06,0,-0.0435,-0.7,-0.1\}$.}}}
\end{figure}
%------------------------------------------------------------

One can find the five `bounding curves', which are depicted by $L(K)$ in Fig.~\ref{fig:gravastarL}, by studying the factorised discriminants of Eqs.~(\ref{DECLn2})-(\ref{DECLp2}), where (\ref{DECLn2})-(\ref{DECLp2}) are viewed as polynomials in $A$, and the restrictions on the parameters given by (\ref{Lr}) that ensure $\rho$ and $P$ are real. The `bounding curves' are the relevant parametric curves which bound the regions of the number of intersections of $\rho(P)$ with $\mathcal{\partial M}$. They are given by,
\bea
L&=&-1/9\,,
\nn
\\
L&=&-6K\,,
\nn
\\
L&=&2K\left(\frac{\sqrt{600K}}{3}-3\right)\,,
\nn
\\
L&=&2K(\sqrt{72K}-3)\,,
\label{boundingcurvesL}
\eea
the ninth order polynomial in $L$, namely $L_9(K)$, which is found by factorising the discriminant of (\ref{DECLn2}). (The actual expression is too large to fit here.) 

There is a small region of the parameter space for $K$ and $L$, with stable gravastar solutions when the interior is a de Sitter space, while there exists an infinitely large region of the parameter space for $K$ and $L$ when the interior is anti--de Sitter space. The region with a Schwarzschild--de Sitter exterior is much larger than the region with a Schwarzschild--anti--de Sitter exterior. The most important bounds on $K$ and $L$, governing the existence of gravastar solutions, are given by $-1/9<L<L_9(K)$.

\begin{figure}[h]
\vbox{\centerline{\scalebox{0.50}{\includegraphics[angle=0]{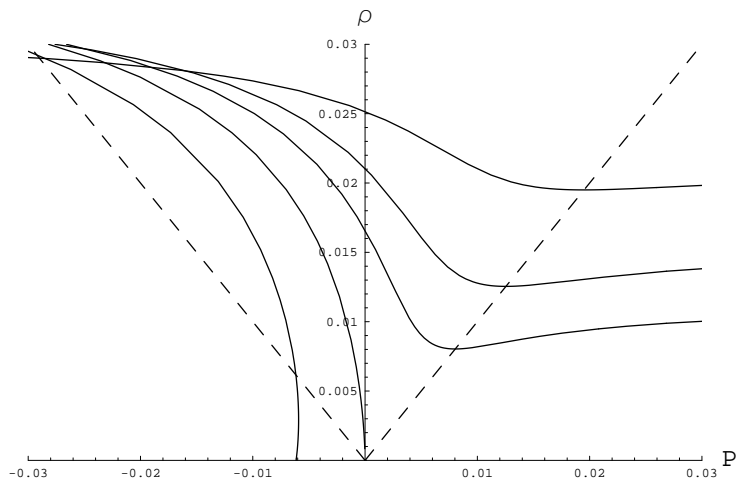}}}
\caption{\label{fig:L0}
{\sl  The behaviour of $\rho(P)$ for the thin shell of a gravastar with $K=0$ and a range of values of $L$. From left to right, $L=\{0.07,0,-0.0435,-0.07,-0.1\}$.}}}
\end{figure}

\begin{figure}[h]
\vbox{\centerline{\scalebox{0.50}{\includegraphics[angle=0]{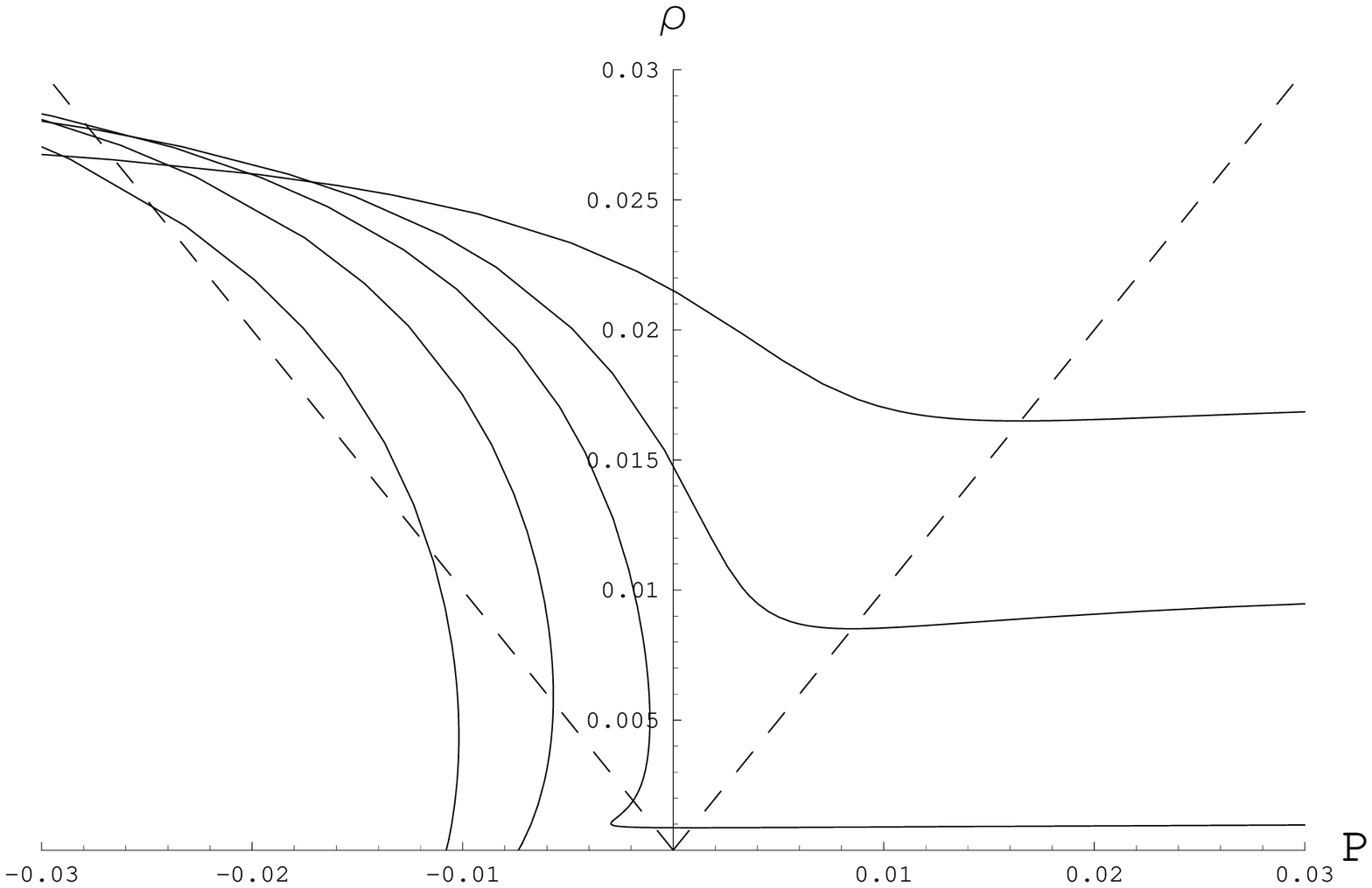}}}
\caption{\label{fig:L001}
{\sl  The behaviour of $\rho(P)$ for the thin shell of a gravastar with $K=0.01$ and a range of values of $L$. From left to right, $L=\{0.07,0,-0.0435,-0.07,-0.1\}$.}}}
\end{figure}

There are different qualitative behaviors that $\rho(P)$ can take depending on the value of the parameters $K$ and $L$, which we present in Figs.~(\ref{fig:Ln001})--(\ref{fig:L001}). The various graphs show the quantitative behaviour of $\rho(P)$ for specific values of $K$ and $L$, but each graph is indicative of the qualitative behaviour of $\rho(P)$ for a particular range of values of $K$ and $L$. The equation of state exhibits a smooth transition from one type of qualitative behaviour to another as the values of $K$ and $L$ pass over the `bounding curves'.

%------------------------------------------------------------
%------------------------------------------------------------
%------------------------------------------------------------
%------------------------------------------------------------

\section{Reissner--Nordstr\"{o}m Gravastar}
%------------------------------------------------------------
\begin{figure}[htb]
\vbox{\centerline{\scalebox{0.33}{\includegraphics[angle=0]{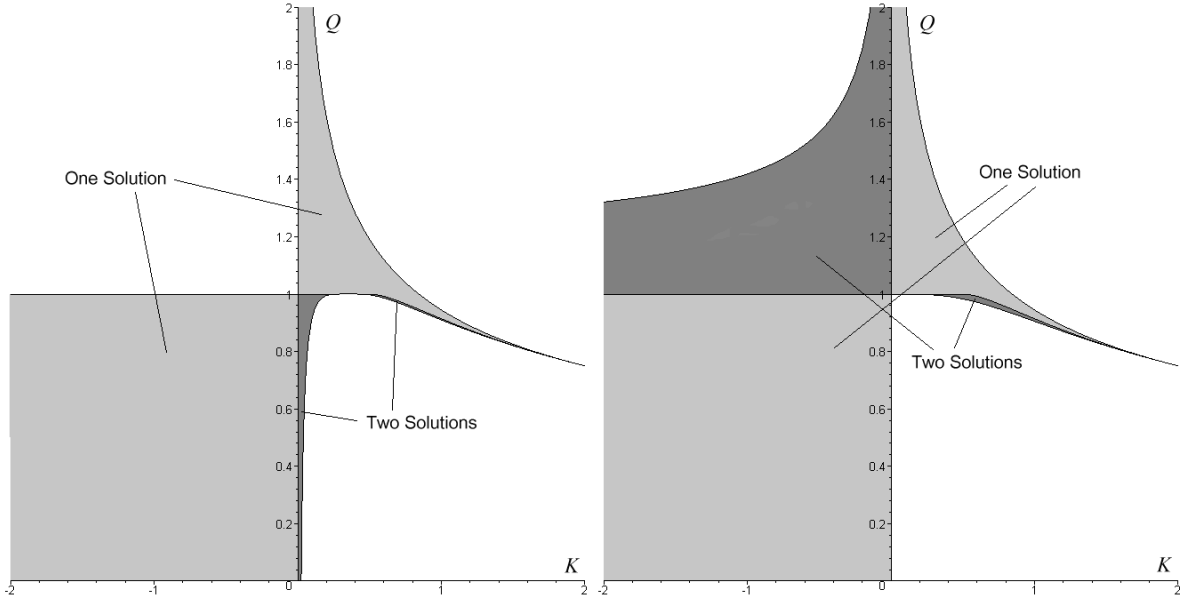}}}
\caption{\label{fig:gravastarCpm}
{\sl  The solution space for stable Gravastars with (A)dS interior and Reissner--Nordstr\"{o}m exterior, satisfying $\rho+P=0$ in the left panel, or $\rho-P=0$ in the right panel. The number of solutions describes how many times $\rho(P)$ passes through the left or right side of $\mathcal{\partial M}$ in the left or right panel respectively. If there are no solutions, then $\rho(P)$ does not cross $\mathcal{\partial M}$, and no interval of $\rho(P)$ satisfies the dominant energy condition.}}}
\end{figure}
%------------------------------------------------------------

\begin{figure}[h]
\vbox{\centerline{\scalebox{0.33}{\includegraphics[angle=0]{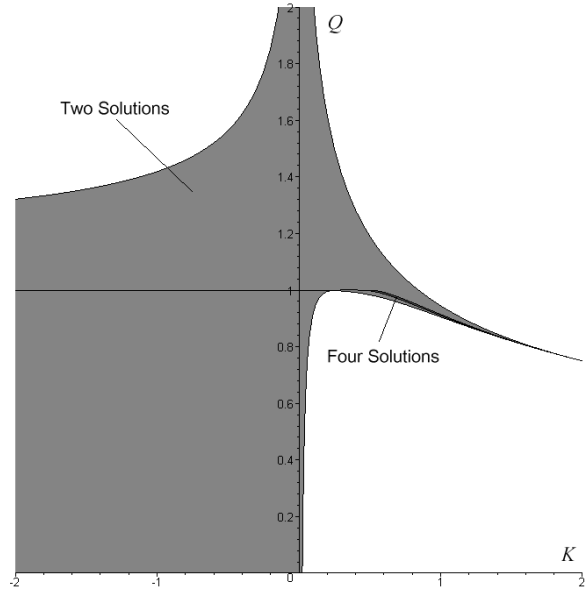}}}
\caption{\label{fig:gravastarC}
{\sl  The solution space for stable Gravastars with (A)dS interior and Reissner--Nordstr\"{o}m exterior, where $\rho(P)$ for the thin shell satisfies either $\rho+P=0 $ or $ \rho-P=0$. The regions with solutions have at least one interval of $\rho(P)$ which lies in $\mathcal{M}$. Therefore, for these regions, there exist stable gravastar solutions. Conversely, the regions without solutions do not allow stable gravastar solutions where the thin shell satisfies the dominant energy condition.}}}
\end{figure}

\begin{figure}[htb]
\vbox{\centerline{\scalebox{0.50}{\includegraphics[angle=0]{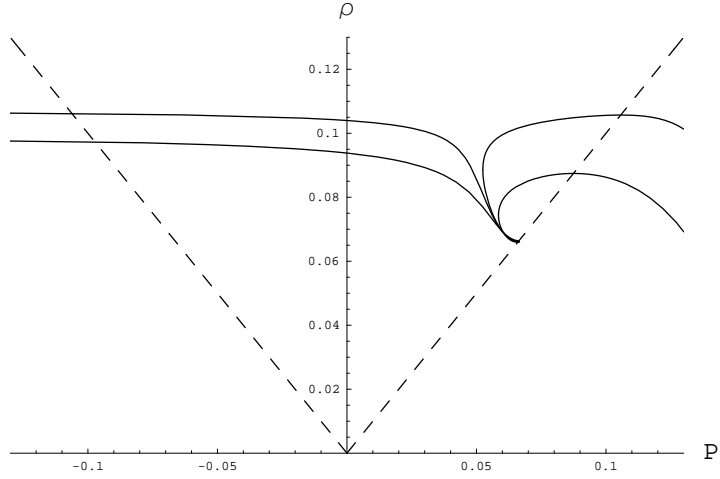}}}
\caption{\label{fig:Qn05}
{\sl  The behaviour of $\rho(P)$ for the thin shell of a gravastar with $K=-0.5$ and a range of values of $Q$. From left to right, $Q=\{0.9,0.99,1.01,1.1\}$. The behaviour for $Q<1$ is qualitatively different to that of $Q>1$.}}}
\end{figure}

\begin{figure}[htb]
\vbox{\centerline{\scalebox{0.50}{\includegraphics[angle=0]{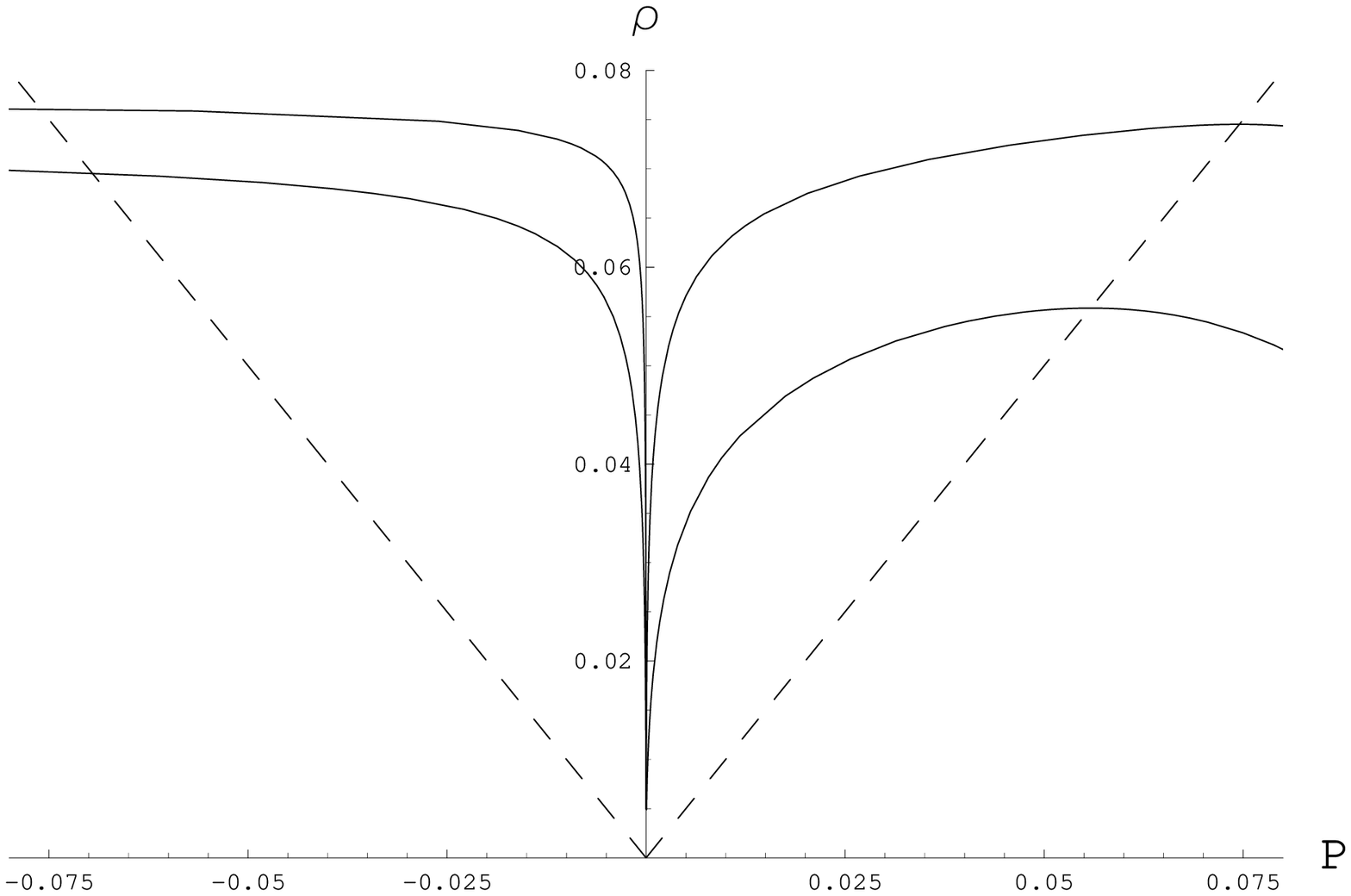}}}
\caption{\label{fig:Q0}
{\sl The behaviour of $\rho(P)$ for the thin shell of a gravastar with $K=0$ and a range of values of $Q$. From left to right, $Q=\{0.99,0.999,1.01,1.1\}$. The behaviour for $Q<1$ is qualitatively different to that of $Q>1$.}}}
\end{figure}

\begin{figure}[htb]
\vbox{\centerline{\scalebox{0.50}{\includegraphics[angle=0]{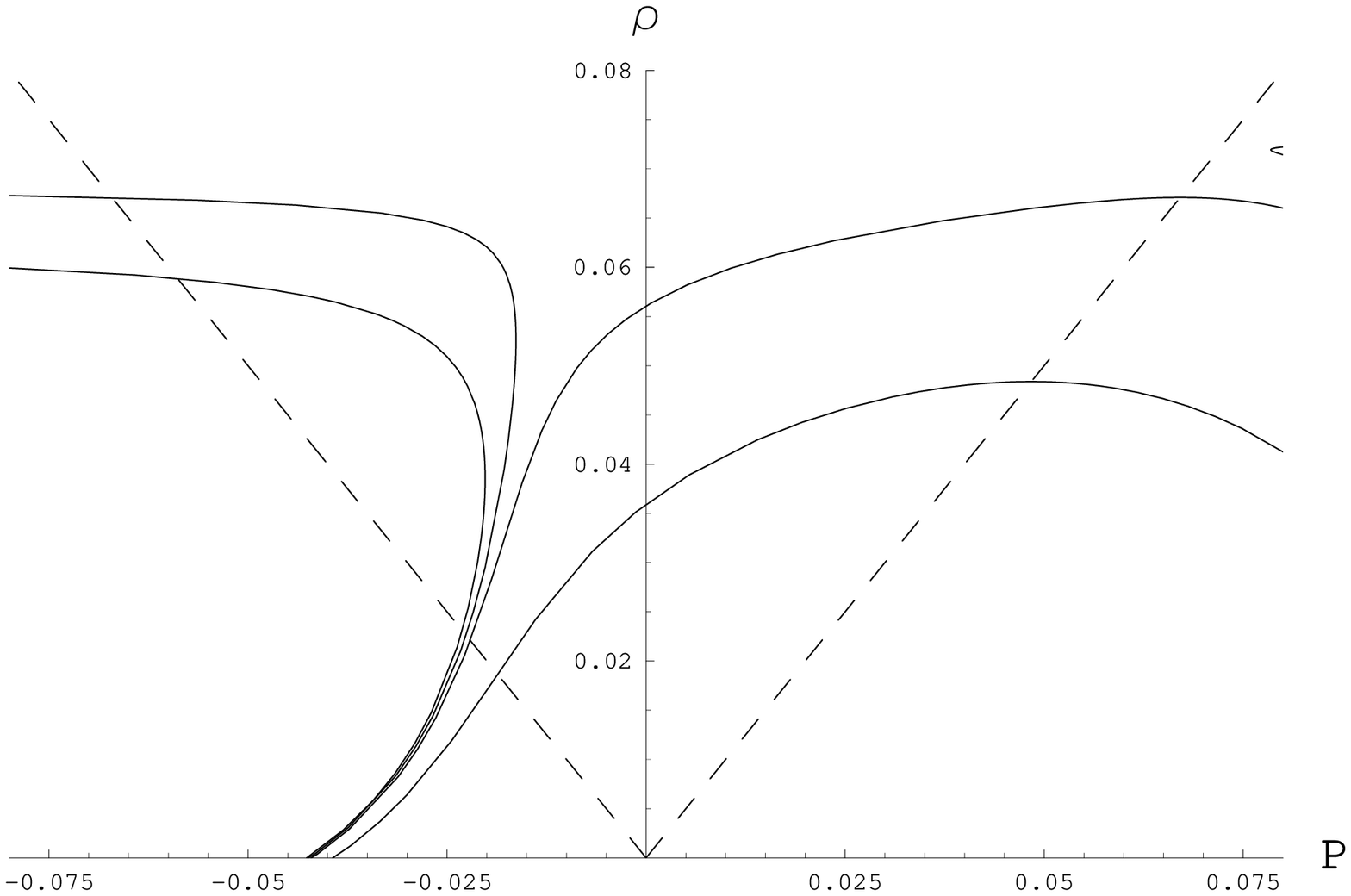}}}
\caption{\label{fig:Q01}
{\sl The behaviour of $\rho(P)$ for the thin shell of a gravastar with $K=0.1$ and a range of values of $Q$. From left to right, $Q=\{0.9,0.99,1.01,1.1\}$. The behaviour for $Q<1$ is qualitatively different to that of $Q>1$.}}}
\end{figure}

\begin{figure}[htb]
\vbox{\centerline{\scalebox{0.50}{\includegraphics[angle=0]{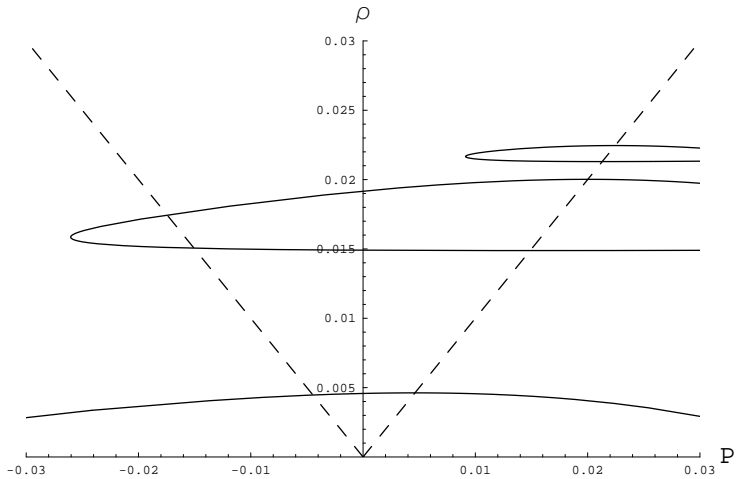}}}
\caption{\label{fig:Q08}
{\sl This figure shows the behaviour of $\rho(P)$ for the thin shell of a gravastar with $K=0.8$ and a range of values of $Q$. From top to bottom, $Q=\{0.945,0.951,1\}$. The qualitative behaviour of the equation of state is continuous over $Q=1$.}}}
\end{figure}

We now consider a gravastar with charge $q$ such that the radial electric field is given by $E_r=\frac{q}{r^2}$. In this case,
\bea
m_+&=&M-\frac{q^2}{2A}\,,
\nn
\\
m_-&=&kA^3\,.
\eea
Following the same procedure as used for the Schwarzschild (A)dS case we find the dimensionless equations that define the endpoints of the interval of $\rho(P)$ in $\mathcal{M}$, located on the left $(\rho+P=0)$ and right $(\rho-P=0)$ sides of $\mathcal{\partial M}$, given respectively by,
\beq
\left(3-\frac{5}{A}+\frac{2Q}{A^2}\right)\sqrt{1-2KA^2} = (3-8KA^2)\sqrt{1-\frac{2}{A}+\frac{Q}{A^2}}\,,
\label{DECQn1}
\eeq
and
\beq
\left(1-\frac{3}{A}+\frac{2Q}{A^2}\right)\sqrt{1-2KA^2} = \sqrt{1-\frac{2}{A}+\frac{Q}{A^2}}\,,
\label{DECQp1}
\eeq
where $Q=\frac{q^2}{M^2}\geq0$. The other dimensionless quantities are as shown in (\ref{dimensionless}). These can be manipulated to give
\bea
0&=&-64 K^2 A^8+ 128 K^2 A^7+(30 K-64 Q K^2) A^6-36 K A^5 -20 Q A+4 Q^2
\nn
\\
&\ &+(24 K Q-50 K) A^4+(40 K Q-12) A^3+(3 Q+25-8 K Q^2)A^2\,,
\label{DECQn2}
\eea
and, 
\bea
0&=&-2 K A^6+12 K A^5+(-8 K Q-18 K) A^4+(24 K Q-4) A^3
\nn
\\
&\ &+(3 Q+9-8 K Q^2) A^2-12 Q A +4 Q^2,
\label{DECQp2}
\eea
respectively. Once again we demand that $\rho$ and $P$ be real which leads to 
\bea
K<\frac{1}{2A^2}
\nn
\\
Q>2A-A^2.
\label{Qr}
\eea
As before, squaring (\ref{DECQn1}) and (\ref{DECQp1}) can lead to an inclusion of a relative minus sign in (\ref{DECQn2}) and (\ref{DECQp2}) respectively, which we account for by placing extra restrictions on $K$ and $Q$ so that that (\ref{DECQn2}) and (\ref{DECQp2}) are consistent with (\ref{DECQn1}) and (\ref{DECQp1}). 
We present our findings for the left and right sides of $\mathcal{\partial M}$, $\rho+P=0$ and $\rho-P=0$, separately in the left and right panels of Fig.~\ref{fig:gravastarCpm}, while we combine the panels in Fig.~\ref{fig:gravastarC}.
%------------------------------------------------------------
%------------------------------------------------------------
Once again the parametric bounds are given by some of the factors of the discriminants of (\ref{DECQn2}) and (\ref{DECQp2}) when they are viewed as polynomials in $A$, the condition that $\rho \geq 0$, and the inequalities in (\ref{Qr}). The case studied by Visser and Wiltshire \cite{Visser:2003ge} is recovered here with $Q=0$. 
There are five bounding curves. Three of them are given by, 
\bea
Q&=&1\,,
\\
Q&=&\frac{3(2 K^2)^{1/3}}{4K}\,,
\\
Q&=&\frac{-1+2\sqrt{2K}}{2K}\,. 
\eea
The fourth one is given by polynomial which is twelfth order in $Q$, found by factorising the discriminant of (\ref{DECQn1}). The fifth bound is given by the octic in $Q$ found by factorising the discriminant of (\ref{DECQp1}).
The qualitative behaviors of the equation of state depend on the choice of the parameters $\{K,Q\}$, as seen in Figs.~\ref{fig:Qn05}--\ref{fig:Q08}.
%------------------------------------------------------------
%------------------------------------------------------------
%------------------------------------------------------------
%------------------------------------------------------------
%------------------------------------------------------------
%------------------------------------------------------------
%------------------------------------------------------------
%------------------------------------------------------------

Extending this section to include a magnetic charge is straight forward, via the substitution $q^2\rightarrow q^2+p^2$ where $p$ is the total magnetic charge of the system. It does not change the results in terms of $Q$, instead it amounts to a redefinition of what we attribute $Q$ to. We note that including a non-zero charge to the exterior of the gravastar greatly increases the range of parameter $K$ of the (A)dS interior for which solutions exist. Gravastars with a charge approximately equal to their mass are `most favoured' for a de Sitter interior in the sense that for such values of $Q$ the least amount of `fine-tuning' of $K$ is required. However, when $Q=1$, $\rho(P)$ appears to be discontinuous for $K<0.423798525400$ (13s.f.).

\section{Discussion}
It has been demonstrated here that the method presented by Visser and Wiltshire \cite{Visser:2003ge} can be used to generate stable gravastars with a non-trivial exterior involving either a vacuum energy or an electromagnetic charge. The Schwarzschild case they studied is retrieved here as a special case. The case with a Schwarzschild-(A)dS exterior, (see Fig.~\ref{fig:gravastarL}), puts bounds on both the interior and exterior values of a vacuum energy allowable for stable gravastar solutions to exist. The most important bound in this case is that the exterior vacuum energy satisfies $-\frac{1}{9}M^{-2}<\Lambda<L_9(K)M^{-2}$.

The Reissner--Nordstr\"{o}m solution (Fig.~\ref{fig:gravastarC}) is very interesting as it is physically reasonable to allow massive stellar objects to have a (small) non-zero electric charge. We find that for a de Sitter interior, the range of the allowable vacuum energy smoothly increases as one adds charge, until the charge is close to the limit $q^2=M^2$. At this point the range of the parameter governing the interior vacuum energy becomes greatly enlarged by comparison with the $q=0$ (Schwarzschild) case. These results are important as they demonstrate that there exists a wide range of allowable gravastars with thin shells satisfying the dominant energy condition which are stable to radial perturbations. 

For future research it would be interesting to study the model in \cite{Visser:2003ge} with $V(a) \neq 0$ or $\dot{a} \neq 0$ as these represent a wider class of configurations of potential gravastars. Another avenue for furthering this research would be to consider a gravastar with external Kerr-geometry, although the method provided by \cite{Visser:2003ge} would have to be generalised first to include off--diagonal terms in the metric, and that is likely to be highly non--trivial.

One source of a realistic gravastar interior might be a fundamental scalar field resulting from Kaluza--Klein compactification of extra dimensions \cite{Gibbons:1985ac} (see \cite{Overduin:1998pn} or \cite{Applequist:1987} for a review). Therefore it would be interesting to see what effect a scalar field has on the stability of the external geometry. Finally, investigating the implications of transitions in the equation of state from one type of qualitative behaviour to another is of interest (see Figs.~\ref{fig:Qn05}--\ref{fig:Q01}); we leave this work for a future paper.

\section*{Acknowledgements}
I would like thank Stephanie Hickford, Ishwaree Neupane, Ewan Orr, Matt Visser and David Wiltshire for their careful reading of this manuscript and helpful comments. This work was supported in part by the Marsden fund of the Royal Society of New Zealand.

\end{document}